\documentstyle[12pt]{article}
\newcommand{\newc}{\newcommand}
\newc{\beq}{\begin{equation}}
\newc{\eeq}{\end{equation}}
\newc{\bea}{\begin{eqnarray}}
\newc{\eea}{\end{eqnarray}}
\newc{\mpl}{M_P}
\newc{\mgrav}{m_{3/2}}
\newc{\mgut}{M_{\rm G}}
\newc{\mstring}{M_{\rm string}}
\newc{\mw}{m_W}
\newc{\msusy}{M_{\rm SUSY}}
\newc{\mint}{M_{\rm int}}
\newc{\gev}{\,\mbox{GeV}}
\newc{\tr}{\mbox{Tr}\,}

\newc{\Si}{\Sigma}
\newc{\eps}{\epsilon}
\newc{\ie}{{\it i.e.\/}}
\newc{\eg}{{\it e.g.\/}}
\def\vev#1{\left\langle #1 \right\rangle}
\newc{\CO}{{\cal O}}
\newc{\gsim}{\lower.7ex\hbox{$\;\stackrel{\textstyle>}{\sim}\;$}}
\newc{\lsim}{\lower.7ex\hbox{$\;\stackrel{\textstyle<}{\sim}\;$}}
\def\NPB#1#2#3{Nucl. Phys. {\bf B#1}, #3 (19#2)}
\def\PLB#1#2#3{Phys. Lett. {\bf B#1}, #3 (19#2)}
\def\PLBold#1#2#3{Phys. Lett. {\bf#1B}, #3 (19#2)}
\def\PRD#1#2#3{Phys. Rev. {\bf D#1}, #3 (19#2)}
\def\PRL#1#2#3{Phys. Rev. Lett. {\bf#1}, #3 (19#2)}
\def\PRep#1#2#3{Phys. Rep. {\bf#1}, #3 (19#2)}

\def\MPLA#1#2#3{Mod. Phys. Lett. {\bf A#1}, #3 (19#2)}

\begin{document}

\begin{titlepage}
\begin{flushright}
{\rm
IASSNS-HEP-98-40\\
RU-98-15\\
{\tt hep-ph/9805240}\\
May 1998\\
}
\end{flushright}
\vskip 2cm
\begin{center}
{\Large\bf Supergravity Resolution of  \\
the Unification to Planck Scale Hierarchy}
\vskip 1cm
{\large
Christopher Kolda${}^1$
and Nir Polonsky${}^2$\\}
\vskip 0.4cm
{\small\it ${}^1$ School of Natural Sciences, Institute for Advanced Study,
Princeton, NJ 08540, USA\\
${}^2$ Department of Physics and Astronomy, Rutgers University, Piscataway,
NJ 08854, USA\\
}
\end{center}
\vskip .5cm
\begin{abstract}
It is demonstrated how the hierarchy between the gauge coupling unification 
scale of minimal supersymmetry and the
Planck (or string)
scale, which resembles in order of magnitude a loop factor, can actually be
explained as such in supergravity-coupled supersymmetry.
A gauge and global singlet field acquires a linear potential
term due to its one-loop supergravity interactions and slides
to the desired scale.
The singlet field can then provide the seed for the breaking of the
unified theory at the appropriate scale via its couplings to fields in
the adjoint representation.

\end{abstract}
\end{titlepage}
\setcounter{footnote}{0}
\setcounter{page}{1}
\setcounter{section}{0}
\setcounter{subsection}{0}
\setcounter{subsubsection}{0}


The minimal supersymmetric extension of the Standard Model (the MSSM)
of electroweask and strong interactions is well known to be consistent
with the unification of the electroweak and strong couplings
at a scale $M_{G} \simeq 3 \times 10^{16}$ GeV.
Given the current measured values of the couplings, the unification
holds at the percentile level (in the units of the unified coupling
$\alpha_{G}\sim 0.04$) with only ${\cal{O}}(1)$ ambiguity
in the unification scale 
(for example, see Refs.~\cite{LP,CPP}). And because the apparent
unification scale is well below the Planck scale, 
Planck-suppressed corrections are sufficiently small that one can trust
the field theory calculation~\cite{LP}. While an impressive result, 
this particular scale for unification is poorly understood.

Specifically, the unification scale lies two orders of magnitude below
the (reduced) Planck scale, $M_{P}$, and an order of magnitude
below the predicted unification scale in perturbative heterotic string
theory, $\sim 5 \times 10^{17}$ GeV \cite{K}.
Within the context of string theory, there are a number of proposals for 
alleviating this discrepancy, including extra matter at intermediate
scales, altered unification conditions, and non-perturbative/M-theory
effects (for a review, see Ref.~\cite{D}.)
Alternatively, there may be a true grand-unified theory (GUT)
in the decades between the
Planck (or string) scale and the phenomenologically determined
unification scale\footnote{
The actual embedding of a unified theory into a string theory
is not straightforward and will not be addressed here. However recent results
on non-perturbative solutions to string theories (\eg, F-theory~\cite{KL})
show, in principle, 
tremendous flexibility in the embeddings that can be arranged.}.

However, in this latter case, one is usually forced to introduce the
scale of GUT-breaking (\ie, the unification scale) as an additional
fundamental scale in the problem. It would clearly be preferable to 
find some mechanism by which one or more of the seemingly fundamental
scales in the theory (the Planck scale, the GUT scale, the supersymmetry-breaking
scale) could be derived from the others. There are already well-motivated
explanations of the supersymmetry-breaking scale as the strong
coupling scale of some
new gauge interaction, replacing it as a fundamental scale in favor
of the Planck scale and an $\CO(1)$ gauge coupling~\cite{witten1}. In this
paper we will derive a mechanism by which the GUT scale can in turn be
extracted as a function of the Planck scale, once supersymmetry is broken.

Several models already exist in the literature for doing just this. One of the
earliest is the
``inverted hierarchy'' model of Witten~\cite{witten2}. In this toy model the
only fundamental scale is the scale of supersymmetry-breaking. The GUT 
is broken at tree-level at a scale determined by the vacuum
expectation value (vev) of a gauge singlet.
That singlet, however, is undetermined at tree-level and only later fixed
by logarithmically divergent corrections to the potential. Because of the
logarithms, the GUT scale is exponentially far from the supersymmetry-breaking
scale. Other more realistic models have also relied on logarithmically
divergent contributions to generate the GUT scale, but down 
from the Planck scale instead of up from the supersymmetry-breaking scale. 
Since the GUT scale is so close to the
Planck scale, though, the exponential hierarchy must be arranged to be small.
The model of Goldberg~\cite{goldberg} generates the GUT scale through the
vev of a singlet when its supersymmetry-breaking mass-squared is
driven negative in the infrared
by large Yukawa interactions (see below). 
A very different model by Cheng~\cite{cheng}
generates the GUT scale as the scale of strong gauge dynamics, that is,
through the usual dimensional transmutation.

Unlike all of these models, we will explain the GUT-to-Planck scale ratio
not in terms of an exponential hierarchy as is generated by logarithmic
corrections, but rather by a loop-factor hierarchy generated by 
quadratic divergences. Such a hierarchy 
can be realized within perturbation theory once a globally
supersymmetric theory is coupled to spontaneously broken
supergravity. (We will assume for concreteness
that supersymmetry-breaking is communicated from
a hidden sector of the theory to the visible/Standard Model sector via
supergravity interactions alone, but will comment later on other 
possibilities.)
We will show that by replacing
dimensionful parameters which correspond to the unification scale
with appropriate couplings of adjoints fields to a singlet,
and  properly treating the supergravity
interactions of the singlet, the desired hierarchy emerges naturally
and is indeed given by a loop factor. The relationship of this mechanism
to other recently proposed scenarios~\cite{np,kpp} using one-loop
supergravity-induced potentials for singlet fields coupled to
fundamentals, rather than adjoints, will be discussed in detail below.

Consider for concreteness an SU(5) theory.
The scale in which the unified symmetry is broken is described
most economically by a mass parameter, $M \simeq \mgut$.
The  minimal choice of a superpotential is:
\beq
W=M\tr\Sigma^2+\lambda\tr\Sigma^3,
\label{eq:su5}
\eeq
where $\Sigma$ is in the adjoint representation of
$SU(5)$. The scalar potential corresponding to Eq.~(\ref{eq:su5}) above
has three degenerate minima at which SU(5) is alternatively unbroken,
broken to SU(4)$\times$U(1), or to SU(3)$\times$SU(2)$\times$U(1). In the
latter two cases, $\Sigma$ receives a vev 
$\sim M/\lambda$, defining the GUT scale.

At this level, the new scale $M=\mgut$ is {\it ad hoc} --- it bears no obvious
relation to any other scale in the theory. It would seem natural to replace
the explicit mass term with a Yukawa interaction, 
$M\tr\Sigma^2 \rightarrow S\tr\Sigma^2$  with $S$ a gauge singlet,
provided that the vev
of $S$ is specified, $\langle S \rangle = \mgut$.
However, the additional equations of motion corresponding to $F_S=0$
drive $\Sigma$ to the origin, $\vev{\Sigma}=0$, leaving SU(5) unbroken.
Satisfactory models involve
at least two distinct Higgs fields in the adjoint of SU(5), $\Sigma_{1,2}$.
For example, consider the  superpotential \cite{goldberg},
\beq
W=\lambda S\tr\Sigma_1\Sigma_2+\lambda'\tr\Sigma_1^2\Sigma_2,
\label{W}
\eeq
where $S$ is again a gauge singlet.
Such a superpotential is the most general one allowed by a combination
of an $R$/phase-symmetry and a $Z_4$ discrete symmetry. In particular,
$S^n$ and all mass terms are forbidden in $W$.
This superpotential allows for SU(5) to break once the singlet $S$
develops a vacuum expectation value,
\bea
F_{\Si_2}=0\Rightarrow\Sigma_1&=&
\frac{\lambda}{\lambda'}S\,\times\,\mbox{diag}(2,2,2,-3,-3)
\nonumber\\
F_{\Si_1}=F_S=0\Rightarrow\Sigma_2&=&0.
\label{minimum}
\eea
However, Eq.~(\ref{W}) alone leaves $S$ undetermined.
As is usually the case in supersymmetric GUT's, the minimum of 
Eq.~(\ref{minimum}) is simply one of several degenerate
vacua, whose degeneracy is lifted in a model-dependent way
once the explicit soft supersymmetry breaking effects
are taken into account. We will assume that it is  lifted such that
the vacuum corresponds to the Standard Model
SU(3)$\times$SU(2)$\times$U(1) configuration.
Eq. (\ref{minimum}) is then stable up to corrections $\sim\mgrav$
\cite{HLW}.

Since $S$ is undetermined in the supersymmetric limit, 
its value must be fixed by its supergravity interactions
once (local) supersymmetry is broken spontaneously in some
hidden sector of the theory. It was proposed~\cite{goldberg,flipped} that a
supergravity-generated scalar-potential
of the form $m_{3/2}^{2}|S|^{2}$, where $m_{3/2}\sim m_W$ 
is the gravitino mass,
is sufficient for that purpose. Because the Yukawa interaction
of Eq.~(\ref{W}) renormalizes the singlet wave-function,
the soft supersymmetry-breaking mass-squared for $S$
diminishes logarithmically with the momentum scale.
Depending on the coupling strength and the relevant
group theory 
factors, at some scale $Q_{0}$ the renormalized singlet mass-squared
may turn negative.
It is straightforward to show that the minimization of the one-loop
effective potential in this case gives $\langle S \rangle \simeq
Q_{0}$~\cite{crossing}.
By carefully choosing the matter content of the theory and the couplings
one may arrange for $Q_{0} \simeq M_{G}$~\cite{goldberg,flipped}.
While possible, such a solution is far from unique and  does not
explain the ``loop-factor''-like hierarchy.
It also implicitly assumes that
the global symmetries of the superpotential are exact symmetries of the vacuum.
Note that the above proposal implies, as one often finds
in supergravity models, a light ($\sim m_{3/2}$) pseudo-Goldstone boson
which carries a large energy density,
which may be cosmologically inconvenient~\cite{cosmology}.

It is likely, however, that the global
symmetries of the superpotential in Eq.~(\ref{W}) are only accidental
symmetries and are due to, \eg, symmetries of the underlying theory
and the renormalizability condition~\cite{global}.
These accidental symmetries of the superpotential may be explicitly
broken in the low-energy theory by Planck-suppressed operators.
Specifically, non-holomorphic operators which violate the symmetries
generically appear in the K\"ahler potential~\cite{BD}. 
The singlet $S$ does not carry
in this case any conserved (gauge or global) quantum numbers.
Being a true singlet, $S$ would be generically
dressed  by quadratically divergent (supergravity) tadpole loop-diagrams
which lead, once supersymmetry is spontaneously broken,
to a linear shift in the effective scalar 
potential~\cite{tadpole,np,kpp}:
\beq
V \rightarrow V +(\gamma\mgrav^2\mpl S + h.c.)
+(\beta\epsilon\mgrav\mpl F_{S} + h.c.)
\label{shift}
\eeq
where $S = S +\theta \psi_{S} + \theta^{2}F_{S}$,
$\beta$ and $\gamma$ are loop-factors
with arbitrary phases,
and $\epsilon$ is a measure in Planckian units of the vacuum
expectation values of the supersymmtry breaking fields in the hidden
sector (see Ref.~\cite{kpp} for a complete discussion). 
Hereafter we set, for simplicity, $\epsilon = 0$ (but see below).

The resulting scalar potential for $S$ reads
\beq
V(S)=m_S^2|S|^2+(\gamma\mgrav^2\mpl S + h.c.).
\label{VS}
\eeq
Obviously, the potential of Eq.~(\ref{VS}) is bounded from below
if and only if (the renormalized) $m_{S}^{2} > 0$;
without loss of generality, we will identify
$m_{S}^{2} = m_{3/2}^{2}$, which is positive definite.
This is an important difference between this model and that of 
Ref.~\cite{goldberg} in which GUT-breaking was driven by $m_S^2$ at
$\mgut$ becoming
negative through renormalization group effects. If the Yukawa coupling, 
$\lambda$, of $S$ to the adjoints is large, 
$m_S^2$ can indeed be driven negative in the
infrared; we will assume that $m_S^2>0$ at the $\mgut$ scale, which simply
puts a model-dependent upper bound (of $\CO(1)$) on $\lambda$.

The potential, Eq.~(\ref{VS}), is minimized for
\beq
S\simeq-\gamma^\dagger\mpl.
\eeq
The loop factor
\beq
\gamma \simeq N\left(\frac{c}{16\pi^{2}}\right)^{n}
\eeq
is determined by: the arbitrary dimensionless
K\"ahler couplings, $c$; the loop-order at which
the divergent supergravity contributions appear, $n$ (generically $n
=1$ or 2); and the multiplicity of the light states circulating
in the tadpole loops, which is summed in $N$.
Hence, one expects $\gamma$ to be in the range of $10^{-2 \pm 2}$,
which is precisely the scale hierarchy we had hoped to achieve.
Note that the resulting hierarchy is independent of the gravitino mass,
and hence, of the scale of supersymmetry breaking in the hidden sector.
Therefore this mechanism works equally well in models with low supersymmetry-breaking
scales, such as so-called gauge-mediated models, so long as $S$ only gets its mass
via supergravity interactions. However, there will always remain in the light 
spectrum a scalar with mass $\sim\mgrav$. 

If we remove our assumption that $\epsilon=0$ and instead
allow $\epsilon \sim 1$, we find that the potential is shifted at its minima
by $V\sim
m_{3/2}^2\mgut^2$, which is the same order as the potential with $\epsilon=0$;
equivalently, ${\Si_2}$ shifts by $\sim\mgrav$. Since we are not
fully analyzing the relative structure of the local minima to an
accuracy better than $\mgrav^2\mgut^2$, this shift is irrelevant at our
current level of discussion. Effective potential corrections
can also be shown to be negligible.

However, in order for this mechanism to work, 
we must forbid
tree-level mixing of hidden and visible fields of the form $K=
ZZ^{\dagger}S/M_{P} + h.c.$, which could render $\gamma \sim 1$, rather
than $10^{-2}$. While we have no symmetry argument for excluding such
terms (unlike for terms linear in $Z$ or $Z^\dagger$), 
we do know that other related terms, such as $ZZ^\dagger QQ^\dagger$ 
(for $Q$ a matter superfield) must have coefficients $\lsim10^{-3}$
in order to avoid large
flavor changing neutral currents; similar coefficients for the
$ZZ^\dagger S$ operator would render it harmless. This is an issue that
arises in all
models of supergravity-mediated supersymmetry breaking, and this model is no 
exception. Also note that $\epsilon = 0$
can resolve cosmological issues \cite{cosmology}
which are typically associated with light hidden-sector moduli.
These issues resurface here, however,  due to the light singlet with 
$\vev{S}\gg\mgrav$.

Lastly, this model can be contrasted to a recently proposed
scenario\footnote{The distinction from low-energy supersymmetry
breaking scenarios \cite{np} is clearer and stems from the 
value of the gravitino mass and from the absence of a $S^{3}$
superpotential interaction here.}
for solving the $\mu$-problem in supergravity-coupled supersymmetry 
models~\cite{kpp}
in which the singlet coupled to a pair of Higgs fields in the
fundamental and anti-fundamental representations of the underlying 
gauge group. In that model,
there existed a local minimum (not the desired one in which the $\mu$-parameter
was generated) in which the singlet received a vev near $\mgut$, but without
generating a large vev for any other Higgs field. In particular, the structure
of that potential guaranteed that the large singlet vev translated into
large and positive squared-masses for the Higgs fields so that no gauge
symmetries were broken. Here, the Higgs fields are in the adjoint
representation so that their interactions need not be vector-like (adjoints are
automatically vector-like with respect to the gauge symmetry, though not
necessarily with respect to the global symmetries). In addition,
their vev's do not have to be aligned in order to cancel $D$-terms,
which vanish automatically for adjoint fields.
Thus a non-trivial superpotential
containing cubic interactions, \ie, $\lambda'\tr\Sigma_{1}^2\Sigma_2$, 
can be devised in order to communicate  the large singlet vev 
to a single self-adjoint field. Hence,
a singlet vev $\sim\mgut$ translates into an adjoint-field vev of the same
order. Note that the two mechanisms cannot both be operative in a
model with only one singlet.
It is possible, however,  that in a model with two singlets, the 
$R$-charges (as well as  discrete charges) of the
fields  could be such that one singlet couples only 
to a vector-like pair of fundamentals while the other couples
only to the adjoint fields.

In summary, we have shown that the generic effective potential
of a gauge and global (flat) singlet in supergravity background
is minimized with the singlet sliding to a scale which is distinguished
from the Planck scale (or the relevant cut-off scale)
by only a loop factor. If properly coupled to a grand-unified theory
it can provide the seed for breaking the unified symmetry
at the correct scale.  Thus, supergravity
can naturally explain the specific choice of the unification scale
in terms of the Planck or string scale.
The breaking of the unified symmetry is then intimately related
to the breaking of supersymmetry,
though only the mass of the singlet fields is dependent on the actual
scale of supersymmetry-breaking.

~

{\it Acknowledgments.}~
It is a pleasure to thank Stefan Pokorski for discussions and
collaboration in earlier stages of this work, and
Hans Peter Nilles for conversations.
The work of CK is supported
by Department of Energy contract DE-FG02-90ER40542 and through
the generosity of Helen and Martin Chooljian. 
The work of NP is
supported by National Science Foundation grant PHY-94-23002.
We would also like to thank
the Aspen Center for Physics, where this work was initiated, for its
hospitality.

\end{document}